\RequirePackage{fix-cm}
\documentclass[twocolumn]{svjour3}
\smartqed
\usepackage[utf8]{inputenc}
\usepackage[T1]{fontenc}
\usepackage[hidelinks]{hyperref}
\usepackage{graphicx}
\usepackage{bm}
\usepackage{mathbbol}
\usepackage{braket}
\usepackage{amssymb,amsmath,amsfonts}
\usepackage{cite}
\usepackage[outercaption]{sidecap}
\newcommand{\up}{\uparrow}
\newcommand{\down}{\downarrow}

\begin{document}

\title{Localization in Quantum Walks with a Single Lattice Defect: \\ A Comparative Study}

\author{Laurita I. da S. Teles \and Edgard P. M. Amorim}

\institute{L. I. S. Teles   \at
Departamento de F\'isica, Universidade do Estado de Santa Catarina, 89219-710, Joinville, SC, Brazil
\\ 
\and
           E. P. M. Amorim   \at
Departamento de F\'isica, Universidade do Estado de Santa Catarina, 89219-710, Joinville, SC, Brazil\\
                           \email{edgard.amorim@udesc.br}
}

\date{Received: date / Accepted: date}

\maketitle

\begin{abstract}

We study how a single lattice defect in a discrete time quantum walk affects the return probability of a quantum particle. This defect at the starting position is modeled by a quantum coin that is distinct from the others over the lattice. This coin has a dependence on $\omega$ which quantifies the intensity of the localization. For some sorts of lattice defects, we show how the localization can have a dependence just on $\omega$, and also, the polar $\alpha$ and azimuth $\beta$ angles of the initial qubit by numerical calculations. We propose a lattice defect whose localization has additional dependence on $\beta+\omega$, leading to extra localization profiles. We compare the quantum walks with our lattice defect to the earlier ones, and we discuss their spreading and survival probability. 


\end{abstract}

\section{Introduction} \label{sec.1}

Quantum random walks \cite{aharonov1993quantum, kempe2003quantum} describe a walk of a quantum particle driven by a unitary time evolution. The time evolution operator is formed by a quantum coin and a conditional displacement operator. The quantum coin acts on the internal degree of freedom of the particle (spin-$1/2$ states) taking it to a new superposition of spin states, and then the displacement operator displaces the particle conditioned to its internal state. The particular quantum dynamics of such walks gives room for interference effects to appear, resulting in ballistic behavior, i.e., a spreading that is quadratically faster than their classical counterparts, as well as the entanglement between spin and position \cite{venegas2012quantum}. Since the random aspect of these walks is introduced by the quantum measurement process, we call them here as quantum walks and we consider just their discrete-time version over a regular one-dimensional lattice.

Earlier studies pointed out that quantum walks have potential applications as quantum search algorithms \cite{shenvi2003quantum,portugal2013quantum}, as universal computational primitives \cite{childs2009universal,lovett2010universal}, and they have been demonstrated as a source of a fruitful perspective for modeling a wide range of phenomena \cite{busemeyer2006quantum,engel2007evidence,dimolfetta2016quantum}. Furthermore, the concrete physical realization in many experimental platforms \cite{wang2013physical} makes them a promising route for engineering a quantum computer. There are excellent reviews which provide a bird's eye view in this fascinating topic (see Refs. \cite{kempe2003quantum,venegas2012quantum,wang2013physical, portugal2013quantum}).

Since quantum walks have ballistic spreading, the probability of a quantum particle to return to its starting position drops exponentially over time \cite{konno2010localization}. However, a single phase or lattice defect in this position can generate constructive interference, localizing the quantum state which evolves to a stationary state on double time steps \cite{wojcik2012trapping}. Our main aim here is to understand by using numerical calculations, how different kinds of defects at the starting position change the localization, in the light of previous studies \cite{konno2010localization,wojcik2012trapping}. Moreover, we present another kind of lattice defect, we compare the quantum walks with this defect to those earlier ones through the localization profiles for distinct initial qubits, and we characterize some dynamics properties such as their spreading and survival probability.

\section{Single defect in quantum walks}\label{sec:2}

The quantum walker is a spin$-1/2$ particle over a discrete lattice whose dynamics is unitary in discrete time of steps. This quantum particle has a qubit and its position, respectively, as its internal and external degrees of freedom. Therefore, a starting quantum walk state is $\ket{\Psi(0)}=\ket{\psi_C}\otimes\ket{j}$ and it belongs to Hilbert space $\mathcal{H}=\mathcal{H}_C\otimes\mathcal{H}_P$. The state $\ket{\psi_C}\in\mathcal{H}_C$ is a qubit or coin state spanned by a spin-$1/2$ basis given by $\{\ket{\up}, \ket{\down}\}$. In the Bloch sphere representation, the qubit can be written as
\begin{equation}
\ket{\psi_C}=\cos\left(\frac{\alpha}{2}\right)\ket{\up}+e^{i\beta}\sin\left(\frac{\alpha}{2}\right)\ket{\down},
\label{psi_C}
\end{equation}
with $\alpha\in[0,\pi]$ and $\beta\in[0,2\pi]$ \cite{nielsen2010quantum}. The position state $\ket{j}\in\mathcal{H}_P$ is spanned by $\left\{\dots, \ket{j-1}, \ket{j}, \ket{j+1}, \ldots\right\}$ with $j\in\mathbb{Z}$ being discrete positions on a regular one-dimensional lattice. Throughout this work, the walk always starts at $j=0$. We consider a discrete time evolution of a quantum walk state $\ket{\Psi(t)}$,
\begin{equation}
\ket{\Psi(t)}=U^t\ket{\Psi(0)},
\label{Psi_t}
\end{equation}
by means of a operator $U$ given by
\begin{equation}
U=\sum_j S[C(j)\otimes\ket{j}\bra{j}],
\label{U}
\end{equation}
composed by a quantum coin $C(j)$ and the conditional displacement operator $S$. The quantum coin acts on the spin states, by taking the internal state from a qubit to another one. Here, we employ the following position-dependent quantum coin,
\begin{equation}
C(j)=\frac{1}{\sqrt{2}}
\begin{bmatrix}
e^{i\omega_{\up\!\up}\!\delta_j} & e^{i\omega_{\up\!\down}\!\delta_j}  \\
e^{i\omega_{\down\!\up}\!\delta_j} & -e^{i\omega_{\down\!\down}\!\delta_j}
\end{bmatrix},
\label{qcoin}
\end{equation}
such that $\delta_j\!=\!1$, if $j\!=\!0$ and $0$, otherwise. For instance, this coin furnishes Fourier $F$ and Hadamard $H$ coins for $(\omega_{\up\!\up},\omega_{\up\!\down},\omega_{\down\!\up},\omega_{\down\!\down})=(0,\pi/2,\pi/2,\pi)$, this means
\begin{equation}
F=\dfrac{1}{\sqrt{2}}
\begin{bmatrix}
1 & i\\
i & 1
\end{bmatrix}, \qquad
H=\dfrac{1}{\sqrt{2}}
\begin{bmatrix}
1 & 1\\
1 & -1
\end{bmatrix},
\end{equation}
respectively, for $j=0$ and $j\neq 0$. So, we get a Fourier coin at the origin and the remaining positions have Hadamard coins for all the time steps, i.e., we have a single Fourier defect along a Hadamard walk. By its turn, the conditional displacement operator
\begin{equation}
S=\ket{\up}\bra{\up}\otimes\ket{j+1}\bra{j}+\ket{\down}\bra{\down}\otimes\ket{j-1}\bra{j}, 
\label{S}
\end{equation}
displaces the spin up (down) state from the position $j$ to $j+1$ ($j-1$), entangling spin and position states. Assuming from the Eq. \eqref{psi_C} that the initial spin up and down amplitudes are, respectively, $a(0,0)=\cos(\alpha/2)$ and $b(0,0)=e^{i\beta}\sin(\alpha/2)$, from the $\ket{\Psi(t)}=U\ket{\Psi(t-1)}$ with Eqs. \eqref{U}, \eqref{qcoin} and \eqref{S}, it is not difficult to obtain the following recurrence relations,
\begin{align}
a(j,\!t)=&\frac{e^{i\omega_{\up\!\up}\!\delta_{j\!-\!1}}}{\sqrt{2}}a(j\!-\!1,\!t\!-\!1)+\frac{e^{i\omega_{\up\!\down}\!\delta_{j\!-\!1}}}{\sqrt{2}}b(j\!-\!1,\!t\!-\!1),\nonumber \\
b(j,\!t)=&\frac{e^{i\omega_{\down\!\up}\!\delta_{j\!+\!1}}}{\sqrt{2}}a(j\!+\!1,\!t\!-\!1)-\frac{e^{i\omega_{\down\!\down}\!\delta_{j\!+\!1}}}{\sqrt{2}}b(j\!+\!1,\!t\!-\!1),
\label{recurr}
\end{align}
constrained to $\sum_j|a(j,t)|^2+|b(j,t)|^2=1$ as the condition of normalization along the walk. By using an iterative procedure, we obtain all amplitudes over time and for all positions. They provide us the return probability to the origin by $P_0(\omega,t)=|a(0,t)|^2+|b(0,t)|^2$ over time. In all models considered here, since the defect generates a localized state, trapping the quantum particle, we also refer to this return probability as a localization.  

\section{Localization at the origin}

The introduction of a lattice defect has been revealed interesting effects in quantum walks. For instance, an appropriate choice of defects along the lattice allows us to reflect, trap or localize the quantum walk state \cite{li2013position-defect, ghizoni2019trojan}. Notably, the localization engendered by some kinds of defects at a particular position was investigated in a few works by means of distinct analytical approaches together with numerical calculations. 

Konno studied the localization via path counting approach \cite{konno2010localization}. W\'{o}jcik et al. added a phase defect to a Hadamard coin at the origin, then after solving the recurrence equations for the stationary states, they calculated the overlap between these states with some particular qubits \cite{wojcik2012trapping}. Zhang et al. went further by displacing the phase defect position, and they detached some spreading properties \cite{zhang2014one-dimensional}. Endo et al. compared W\'{o}jcik's and Konno's analytical approaches within this context \cite{endo2014aone-dimensional}.

The defects in quantum walks studied by Konno and W\'{o}jcik can be modeled through the use of different sets of parameters in our position-dependent quantum coin: (i) for $\omega_{\up\!\up,\down\!\down}=0$, $\omega_{\up\!\down}=\omega$, and $\omega_{\down\!\up}=-\omega$ we have Konno coin at the origin \cite{konno2010localization},
\begin{equation}
C_K=\frac{1}{\sqrt{2}}
\begin{bmatrix}
1 & e^{i\omega}  \\
e^{-i\omega} & -1
\end{bmatrix},
\label{Konnocoin}
\end{equation}
(ii) for $\omega_{\up\!\up,\up\!\down,\down\!\up,\down\!\down}=\omega$ we get the phase defect from W\'{o}jcik's work \cite{wojcik2012trapping},
\begin{equation}
C_W=e^{i\omega}H,
\label{Wojcikcoin}
\end{equation}
and (iii) our set of parameters are $\omega_{\up\!\up}=0$, $\omega_{\up\!\down,\down\!\up}=\omega$, and $\omega_{\down\!\down}=2\omega$ resulting in 
\begin{equation}
C_T=\frac{1}{\sqrt{2}}
\begin{bmatrix}
1 & e^{i\omega}  \\
e^{i\omega} & -e^{2i\omega}
\end{bmatrix},
\label{Telescoin}
\end{equation}
which allows us to add a Fourier defect ($\omega=\pi/2$) differently from earlier proposals. Notice that the quantum walks driven by Hadamard and Fourier coins can have distinct transport and entanglement features \cite{orthey2017asymptotic,orthey2019connecting}, the change between them in disordered scenarios enhances the entanglement \cite{vieira2013dynamically,zeng2017discrete,orthey2019weak,pires2019multiple} and localizes a quantum particle \cite{vieira2014entangling,buarque2019aperiodic}.

Konno coin generates the same localization regardless of the initial qubit, once the terms $e^{\pm i\omega}$ impose a relative phase of $\omega$ between two-level states and it implies that the dependence on the qubit $(\alpha,\beta)$ vanishes after calculating $P_0(t)$. For instance, $P_0(\omega,\!4)=\!3/8\!-\!(1/4)\cos\omega$ and the next probabilities get $P_0(\omega,t)=P_0(\omega,t+2)$ for $t=4k$ with $k\in\mathbb{N}^*$, so we can write each probability as 
\begin{equation}
P_0(\omega,\!4k)\!=\!\sum_{n=0}^{2k-1}\!A_n\cos(n\omega), 
\label{expK}
\end{equation}
where $A_n$ correspond to the coefficients of the series. One of the main results of Konno's work \cite{konno2010localization} was an expression which supplied $\lim_{n\rightarrow\infty} P_0(\omega, 2n)$ given by
\begin{equation}
c(\omega)=\left(\frac{2-2\cos\omega}{3-2\cos\omega}\right)^2,   
\label{Konnoeq}    
\end{equation}
and rewriting it as a Fourier cosine series, we have
\begin{align}
F_c(\omega)=&\left(1-\frac{7}{5\sqrt{5}}\right)+\left(2-\frac{26}{5\sqrt{5}}\right)\cos\omega+\nonumber \\
&\left(7-\frac{79}{5\sqrt{5}}\right)\cos(2\omega)+\left(22-\frac{246}{5 \sqrt{5}}\right)\cos(3\omega )\nonumber \\
&+\mathcal{O}(n\geq4),
\label{Konnoexp} 
\end{align}
it is possible to check if our results from Eq. \eqref{expK} converge on the expansion above. 

\begin{figure}[h!]
\center\includegraphics[width=\linewidth]{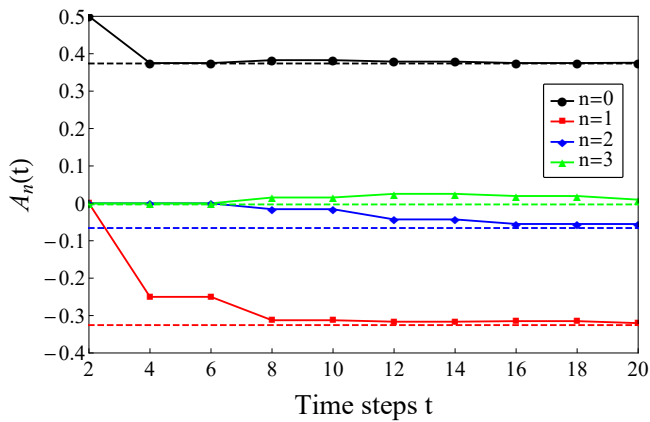}
\caption{Coefficients $A_n$ (solid lines) for $n=0$ (black), $1$ (red), $2$ (blue), and $3$ (green) from Eq. \eqref{expK} and the corresponding coefficients from the Fourier expansion $F_c(\omega)$ (dashed lines) given by Eq. \eqref{Konnoexp}. The lines connecting points are guides for the eyes.}
\label{fig:1}
\end{figure}

Figure \ref{fig:1} shows a plot of the first coefficients $A_n$ of Eq. \eqref{expK} compared to the coefficients of Eq. \eqref{Konnoexp}. Let us consider $\omega=\pi/2$, then $c=4/9\approx0.444$, while for $t=20$, $P_0\approx0.458$ being $\approx0.431$ the sum of the terms with $n=0$ to $3$ or also, for $\omega=\pi$, $c=0.640$, and $P_0\approx0.646$ with $\approx0.631$ corresponding to the same sum above. Although there are tiny differences between Konno's and our results for a few time steps, these differences become more pronounced when $\omega\rightarrow0$ and $2\pi$; therefore, longer quantum walks are required to verify the convergence.

\begin{figure}[h!]
\center\includegraphics[width=\linewidth]{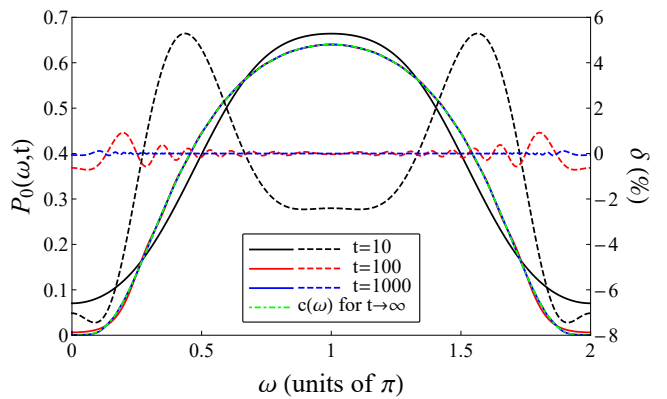}
\caption{Probability $P_0(\omega,t)$ (solid lines) of localization and $\delta=c(\omega)-P_0(\omega,t)$ (dashed lines) after $10$ (black), $100$ (red), and $1000$ (blue) time steps. Plot of $c(\omega)$ (green dot-dashed line) given by Eq. \eqref{Konnoeq} \cite{konno2010localization}.}
\label{fig:2}
\end{figure}

Figure \ref{fig:2} shows the probability $P_0$ as a function of the parameter $\omega$ of Konno coin for quantum walks after different time steps. We see that quantum walks with $100$ and $1000$ time steps can probe the behavior of $t\rightarrow\infty$ with discrepancies $|\delta|$, respectively, smaller than $10^{-2}$ and around $10^{-3}$. Then, from this point our study will focus on showing the probability for the qubit-dependent localization regarding W\'{o}jcik's work and ours by carrying out quantum walks with $1000$ time steps.

W\'{o}jcik deals with a coin $e^{i\omega}H$, a Hadamard coin with a global phase $\omega$. Notice that it keeps the same relative phase between spin states of Hadamard coin. However, this global phase at the origin is taken to other positions breaking the relative phase symmetry of Konno coin by the emergence of a dependence on the initial qubit $(\alpha, \beta)$. For example, $P_0(\omega,\!4)\!=\!3/8\!-\!(1/4)\cos\omega\!-\!(1/4)\sin\alpha\sin\beta\sin\omega$, then the probabilities are
\begin{equation}
P_0(\omega,\!4k)\!=\!\sum_{n=0}^{2k-1}\![B_n\cos(n\omega)\!+\!C_n\sin\alpha\sin\beta\sin(n\omega)], 
\label{expW}
\end{equation}
where $B_n$ and $C_n$ correspond to the coefficients of the series. This expression suggests that the dependence on the initial qubit only appears for the ones in which $\sin\alpha\sin\beta\neq0$.  
\begin{figure}[h!]
\center\includegraphics[width=\linewidth]{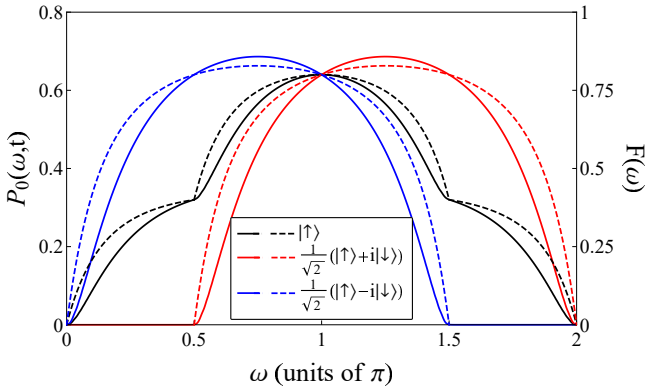}
\caption{Probability $P_0(\omega,t)$ of localization (solid lines) and total overlap $F(\omega)$  (dashed lines) between stationary states and the initial qubits \cite{wojcik2012trapping} given by Eq. \eqref{overlap}. The corresponding qubits $(\alpha,\beta)$ are $(0,0)$ (black), $(\pi/2,\pi/2)$ (red), and $(\pi/2,3\pi/2)$ (blue).}
\label{fig:3}
\end{figure}

Figure \ref{fig:3} shows the probability $P_0$ for three initial qubits together with the total overlap $F(\omega)$ between stationary states and these qubits \cite{wojcik2012trapping}. The overlap is $F_+$ for $\omega\in(0,\pi/2]$, $F_++F_-$ for $\omega\in(\pi/2,3\pi/2)$ and $F_-$ for $\omega\in[3\pi/2,2\pi)$ such that
\begin{equation}
F_{\pm}=|\braket{\Phi_{\pm}|\psi_C}|^2, 
\label{overlap}
\end{equation}
where
\begin{align}
\ket{\Phi_{\pm}}&=\sqrt{\frac{1}{2}+\frac{1}{4\cos\omega\mp4\sin\omega-6}}
\begin{bmatrix}
1 \\
\mp i
\end{bmatrix},
\label{Wojcikstates}
\end{align}
correspond to the stationary states \cite{wojcik2012trapping}. Therefore, the comparison between these two calculations reveals that the stationary method can explore the localization profile introduced by a phase defect at origin \cite{wojcik2012trapping} or other positions \cite{zhang2014one-dimensional} in good agreement with our calculations.

\begin{figure}[h!]
\center\includegraphics[width=\linewidth]{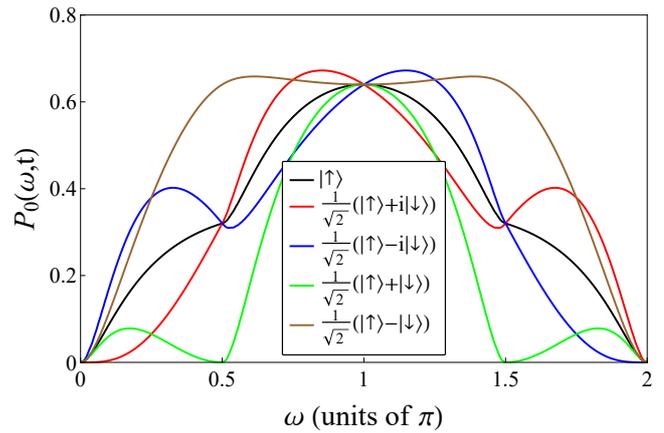}
\caption{Probability $P_0(\omega,t)$ of localization (solid lines) for distinct initial qubits. The qubits $(\alpha,\beta)$ are $(0,0)$ (black), $(\pi/2,\pi/2)$ (red), $(\pi/2,3\pi/2)$ (blue),  $(\pi/2,0)$ (green), and  $(\pi/2,\pi)$ (brown).}
\label{fig:4}
\end{figure}

The phase defect in W\'{o}jcik's work and our lattice defect here have some similarities and differences. The determinants of both coins are the same. However, if we rewrite our coin with a global phase yields
\begin{equation}
C_T=\frac{e^{i\omega}}{\sqrt{2}}
\begin{bmatrix}
e^{-i\omega} & 1  \\
1 & -e^{i\omega}
\end{bmatrix},
\label{gpTelescoin}
\end{equation}
which shows that besides the global phase, this coin introduces an extra relative phase $\omega$ between spin states. Therefore, this results in a translation of $\beta$ by $\omega$ after this coin acts on the qubit. For instance, we obtain $P_0(\omega,\!4)\!=\!3/8\!-\!(1/4)\cos\omega\!-\!(1/4)\sin\alpha\sin(\beta+\omega)\sin\omega$ and the probabilities provide us
\begin{equation}
P_0(\omega,\!4k)\!=\!\sum_{n=0}^{2k-1}\![B_n\cos(n\omega)\!+\!C_n\sin\alpha\sin(\beta+\omega)\sin(n\omega)],
\label{expT}
\end{equation}
and the coefficients of the series $B_n$ and $C_n$ are the same as W\'{o}jcik's case, with a new dependence on $\beta+\omega$, though. This dependence brings us new localization profiles as shown in Fig. \ref{fig:4}. 
\begin{figure}[h!]
\center\includegraphics[width=\linewidth]{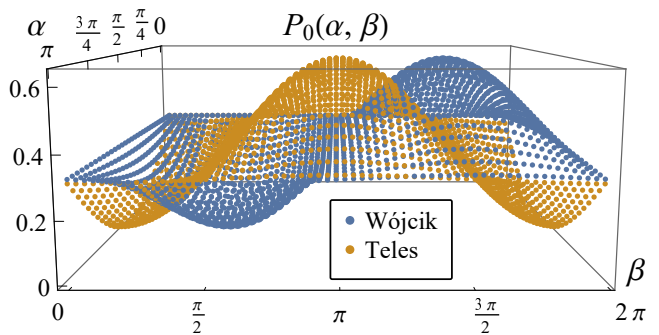}
\caption{Probability $P_0$ for $\omega=\pi/2$ as function of the initial qubit $(\alpha,\beta)$ for the phase defect from W\'{o}jcik's work (blue dots) \cite{wojcik2012trapping} and our proposal of lattice defect (orange dots).}
\label{fig:5}
\end{figure}

The integration of $P_0$ over the whole Bloch sphere (all qubits) leads to the same average localization profile in W\'{o}jcik's work as well as in ours, which corresponds to the qubit $(\alpha,\beta)=(0,0)$. As an example, Fig. \ref{fig:5} shows $P_0$ for $\omega=\pi/2$ of quantum walks starting from many qubits for both cases, i.e., a phase defect of $\pi/2$ and a Fourier defect at origin. We have identical profiles but displaced by $\pi/2$ as expected.

\section{Discussion}

All quantum walks considered here have as a common feature to trap part of the state at origin without preventing the propagation of the remaining non-trapped part. It could be interesting to take a better look into the behavior over time of such walks. First, we contrast their spreading properties to the Hadamard quantum walk. Second, we check the asymptotic dynamic features of the trapped part of the state by comparing it to its initial state.  

\begin{figure}[h!]
\center\includegraphics[width=\linewidth]{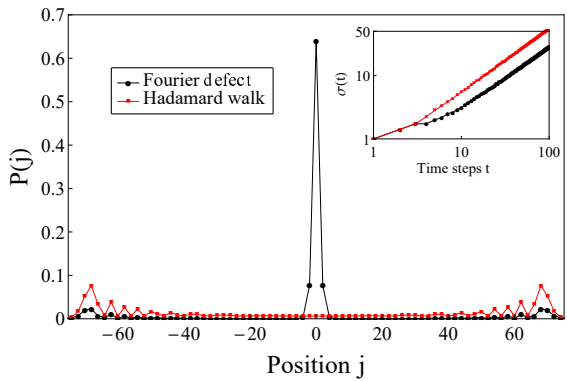}
\caption{Probability distribution $P(j)$ of a quantum walk with a single Fourier defect at origin (black circle) and a Hadamard walk (red square) starting from the qubits, respectively, $(\pi/2,\pi)$ and $(\pi/2,\pi/2)$ after $100$ time steps. Points are plotted only for even positions and the lines connecting them are guides for the eyes. Inset: dispersion $\sigma(t)=(\braket{j^2}-\braket{j}^2)^{\frac{1}{2}}$ for both cases.}
\label{fig:6}
\end{figure}

Let us consider a quantum walk with a single Fourier defect at origin starting from the qubit $\frac{1}{\sqrt{2}}(\ket{\up}-\ket{\down})$. This specific case has the highest localization probability between the qubits studied as shown in Fig. \ref{fig:4}. For sake of comparison, let us take a Hadamard walk starting from the qubit $\frac{1}{\sqrt{2}}(\ket{\up}+i\ket{\down})$. We choose distinct initial states for each of them because they lead to symmetrical probability distributions. Figure \ref{fig:6} shows the probability distribution and dispersion over time. The non-trapped part of the state is given by some symmetrical peaks separated by $\sqrt{2}t$ positions which also occurs to the Hadamard walk \cite{orthey2019connecting}. They exhibit ballistic behavior having quite similar diffusion coefficients (see the inset). 

It is worth mentioning that the localization presented here is remarkably distinct from an Anderson localization. This localization appears when a quantum particle propagates over a disordered medium and it is characterized by the absence of diffusion. One of the ways to give birth to an Anderson localization in the quantum walks context is by means of a static disorder, i.e., when then quantum coins at each position are randomly chosen \cite{vieira2014entangling}. Then, it is reasonable to suppose that a random set of lattice defects over many positions such as those shown here naturally leads to a strong localization ceasing any kinds of diffusion. 

Another way to interpret a single defect is by seeing it as an absorbing boundary. When a classical particle starts a walk in the presence of such boundary, it is totally absorbed. However, it does not happen for a quantum particle. For instance, an initial state $\ket{\up}\otimes\ket{0}$ driven by a Hadamard coin and neighboring an absorbing barrier has an escape probability of $\approx0.36$ \cite{kempe2003quantum,venegas2012quantum}. In particular, this part of the non-absorbed state still keeps a ballistic diffusion. In this sense, there is a considerable resemblance between a quantum walk with a single defect and the one in the presence of an absorbing barrier.

\begin{figure}[h!]
\center\includegraphics[width=\linewidth]{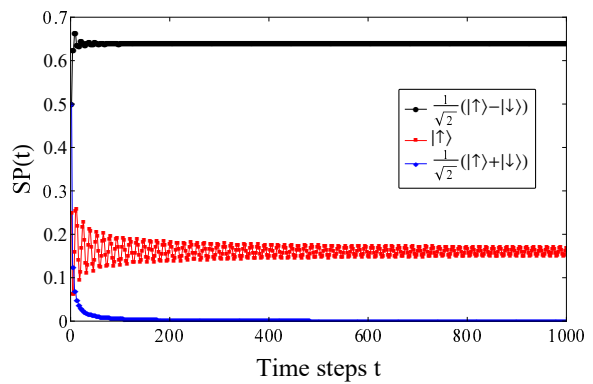}
\caption{Survival probability $SP(t)=|\braket{\Psi_0(t)|\psi_C}|^2$ where $\ket{\Psi_0(t)}=\braket{0|\Psi(t)}$ of quantum walks with a single Fourier defect at origin. The qubits $(\alpha,\beta)$ are $(\pi/2,\pi)$ (black circle), $(0,0)$ (red square), and $(\pi/2,0)$ (blue diamond). The qubits $(\pi/2,\pm\pi/2)$ have the same behavior of $(0,0)$. Points are plotted only for even time steps and the lines connecting them are guides for the eyes.}
\label{fig:7}
\end{figure}

The survival probability can be understood as how much of the initial state remains at the starting position of the walker over time. It can also be interpreted as fidelity to the initial state and it is particularly convenient to analyze the asymptotic dynamics of the state \cite{buarque2019aperiodic, Martin-Vazquez2020Optimizing}. Figure \ref{fig:7} shows the survival probability of the state over time. When the quantum walks starting from the qubits $\frac{1}{\sqrt{2}}(\ket{\up}-\ket{\down})$ and $\frac{1}{\sqrt{2}}(\ket{\up}+\ket{\down})$ the ratio $SP/P_0$ over time is always $1$. Therefore the lattice defect does not change the initial relative phase between spin states in the localization position. However, for the remaining qubits considered here, this ratio converges to $\approx0.5$ indicating a change on the original phase of the qubit. 

\section{Conclusions}

We characterized the localization profiles of quantum walks with three distinct lattice defects at the starting position by carrying out numerical calculations. These localization profiles correspond to the probabilities of a quantum particle to return to its initial position as function of a parameter $\omega$ to quantify the intensity of the localization.

Our main results can be summarized in the following statements: (i) our numerical calculations are in good agreement with Konno's and W\'{o}jcik's models; (ii) while the lattice defect from Konno's work introduces a localization regardless of the initial qubit, the phase defect from W\'{o}jcik's work and our lattice defect are qubit-dependent; (iii) we depicted the dependence on $\omega$ and the initial qubit given by $(\alpha, \beta)$ for each proposal; (iv) we found out that our proposal has a dependence on $\beta+\omega$, distinct from the phase defect; (v) this novel dependence leads to extra localization profiles; (vi) the presence of a single lattice defect does not prevent the ballistic spreading; and (vii) the survival probability indicates that the defect can modify the relative phase of the localization with respect to the initial qubit.

At last, we hope our results can be used to foster the discussion about the localization in such walks and inspire experimental researchers to test our findings.

\section*{Acknowledgements}
This study was financed in part by the Coordena\c{c}\~ao de Aperfei\c{c}oamento de Pessoal de N\'ivel Superior - Brasil (CAPES) - Finance Code 001. The authors thank J. Longo for her careful revision of the manuscript.

\end{document}